\documentclass{article}
\usepackage{graphicx}
\usepackage{epstopdf}
\usepackage{subfig}
\usepackage{amsfonts,amsmath}
\usepackage{cite}
\usepackage{hyperref}
\usepackage{setspace}
\date{}

\begin{document}

\begin{tabular}{ p{\textwidth} }
\begin{center}
\begin{doublespace}
{\LARGE The Escape Problem in a Classical Field Theory With Two Coupled Fields}\\
{\large \bf Lan Gong$^{1}$ and D. L. Stein$^{1,2}$}\\
{\small \tt \href{mailto:lan.gong@nyu.edu}{lan.gong@nyu.edu}} ~ {\small \tt \href{mailto:daniel.stein@nyu.edu}{daniel.stein@nyu.edu}}\\
\end{doublespace}
$^{1}${\small \sl Department\ of Physics,~New York University,~New York,~NY 10003}\\
$^{2}${\small \sl Courant Institute of Mathematical Sciences,~New York University,~New York,~NY 10003}
\end{center}
\end{tabular}

\begin{abstract}
We introduce and analyze a system of two coupled partial differential
equations with external noise. The equations are constructed to model
transitions of monovalent metallic nanowires with non-axisymmetric
intermediate or end states, but also have more general applicability.  They
provide a rare example of a system for which an exact solution of
nonuniform stationary states can be found.  We find a transition in
activation behavior as the interval length on which the fields are defined
is varied.  We discuss several applications to physical problems.
\end{abstract}
\small
\renewcommand{\baselinestretch}{1.25}
\normalsize

\section{Introduction}
\label{sec:intro}

A locally stable system subject to random perturbations will eventually be
driven from its basin of attraction by a sufficiently large fluctuation.
The case of weak noise is particularly important, and often arises in
stochastic modeling of physical and non-physical
systems~\cite{Doering90,Gardiner85,Horsthemke84,Risken89,Brand89,
Schuss80,vanKampen81,McClintock89,Hanggi90,Talkner95}.  Noisy systems in
which spatial variation of some intrinsic property cannot be ignored arise
in numerous contexts and are therefore of particular
interest~\cite{GarciaOjalvo99}.  Examples include micromagnetic domain
reversal~\cite{MSK05,MSK06}, pattern nucleation~\cite{Cross93,Tu97,Bisang98},
transitions in hydrogen-bonded ferroelectrics~\cite{Dikande97}, dislocation motion across Peierls
barriers~\cite{GB97}, and structural
transitions in metallic nanowires~\cite{BSS05,BSS06}.  

Noise-induced escape can occur through either classical or quantum
processes. In classical systems, where escape involves activation over a
barrier, the source of the noise is often, but not always, thermal in
origin. The field-theoretic techniques for computing escape rates in these
systems were developed by Langer~\cite{Langer67,Langer69}, who considered
the homogeneous nucleation of one phase inside another. In a quantum system
at sufficiently low temperature, escape occurs by tunneling through a
barrier; although the process is physically different, the mathematical
formalism is similar to that of the classical case. The basic theory here
was worked out by Coleman and Callan~\cite{Coleman77,CC77,Coleman79} in the
context of quantum tunneling out of a ``false vacuum''.  A review of the
basic approach in both cases can be found in~\cite{Hanggi90,Schulman81}.

All of these consider the case of an infinite system.  The equivalent
problem on a finite spatial domain was first studied by Faris and
Jona-Lasinio~\cite{Faris82,Martinelli89}, who developed a large deviation
theory of the leading-order exponential term in activated barrier crossing
for the special case of Dirichlet boundary conditions.  General techniques
for computing the activation prefactor were developed by
Forman~\cite{Forman87} and McKane and Tarlie~\cite{McKane95}.  In general,
computation of the subdominant behavior of activated processes (the rate
prefactor, the distribution of exit points, and related quantities)
requires knowledge of the transition state, which describes the system
configuration at the col, or barrier `top'.  When this state is spatially
nonuniform, as typically occurs at all but the shortest system sizes, it
can typically only be found as the solution of a nonlinear differential
equation.  This usually requires numerical techniques; nontrivial problems
where analytical solutions can be found are rare.

When spatial variation can be ignored, the dynamical evolution of a noisy
system can be modelled using a stochastic ordinary differential equation;
when it can't, a stochastic partial differential equation is required. This
can and does lead to new phenomena; one of these is a phase transition in
activation behavior as some system parameter is varied.  In classical
systems confined to a finite domain this transition can occur as system
size (or some other parameter such as external magnetic field in a magnetic
system) changes~\cite{MS01b,Stein04}; in quantum systems, an analogous
quantum tunneling$\leftrightarrow$classical activation transition occurs as
temperature
increases~\cite{Affleck81,Wolynes81,Grabert84,Chudnovsky92,Kuznetsov97,BSS08}.
This phase transition has a profound effect on activation behavior, and can
have physically observable consequences~\cite{BSS06}.

Most of the cases studied to date require only a single classical field to
describe the spatial variation of the system. However, situations can arise
in which {\it two\/} (or more) fields are necessary to model the system.
These have received little attention to date; one notable exception is the
analysis of Tarlie~{\it et al.\/}~\cite{Goldbart94} which studied phase
slippage in conventional superconducting rings.  In this case, the spatial
and gauge symmetries imposed by the physics allowed an exact solution of
the transition state to be found.  More generally, such symmetries are
absent, and it is of interest to find more general systems in which exact
solutions can similarly be found.

In this paper we propose and analyze one such model that may have wide
applicability; it can be regarded as a generalization of the system studied
in~\cite{Goldbart94}.  The model was motivated by the growing body of
research on metallic wires of several nanometer diameters and lengths of
the order of tens of
nanometers~\cite{RAV96,ST96,KRJ96,KT97,OKT98,BYR98,YYR99,KT00}, which
represent the ultimate size limit of conductors and are of interest from
the point of view of both fundamental physics and technological
applications. However, as we discuss in the conclusions, the model and its
solutions may apply as well to other problems of interest.  We therefore
confine our discussion in this paper to the mathematical features of the
model and its solution.  Its specific application to nanowires --- which
will necessarily involve additional modelling --- will appear
elsewhere~\cite{BGSSinprep}.

\section{Monovalent Metalic Nanowires}
\label{sec:nanowires}

Nevertheless, to set the stage for introduction of the model, we briefly
discuss the problem of stability and lifetimes of nanowires composed of
metals from either the alkali or the noble metal groups.  Because the
stresses induced by surface tension at the nanometer lengthscale exceeds
the Young's modulus, such wires are subject to deformation under plastic
flow~\cite{KSBG99}. A purely classical wire would therefore be subject to
breakup due to the Rayleigh instability, and this in fact has been observed
for copper nanowires annealed between 400 and 600$^\circ$C~\cite{MBCNT04}.
However, a nanowire is sufficiently small for quantum effects to also play
a role, and indeed a quantum linear stability
analysis~\cite{KSBG99,Kassubek00,KSGG01,SKG01} showed that at discrete
values of the radius, the Rayleigh instability is suppressed. These radii
correspond to conductance ``magic numbers'' that agree with those observed
in experiments~\cite{YYR99,YYR00,YRY01}.

The linear stability analysis, however, ignores thermal noise that can
induce rare but large radius fluctuations leading to breakup. A
self-consistent approach to determining lifetimes~\cite{BSS04,BSS05}, which
modelled thermal fluctuations through a stochastic Ginzburg-Landau
classical field theory, obtained quantitative estimates of alkali nanowire
lifetimes, in good agreement with experimentally inferred
values~\cite{YYR99,YYR00,YRY01}. The theory restricted itself to perfectly
cylindrical wires, so that a single classical field could be used to
represent radius fluctuations along the length of the wire.

To test the assumption of axial symmetry along the cylinder axis,
Urban~{\it et al.\/}~\cite{UBZSG04} performed a stability analysis of metal
nanowires subject to non-axisymmetric perturbations. They were able to show
that, at certain mean radii and aspect ratios, Jahn-Teller deformations
breaking cylindrical symmetry can be energetically favorable, leading to
stable nanowires with non-axisymmetric cross sections. They predicted that
a typical mechanically controllable break junction experiment should
observe roughly 3/4 cylindrical wires and 1/4 noncylindrical.

The mathematical problem can be understood as follows. Urban~{\it et
al.\/}~\cite{UBSG06} considered non-axisymmetric deformations in the wire
cross-section of the form $\cos(m\phi)$, i.e., having $m$-fold symmetry.
The radius function describing the surface of a wire of cross-sectional
area $\pi\rho(z)^2$ at position $z$ along the wire length is then
\begin{equation}
\label{eq:deformation}
R(\rho,\phi)=\rho(z)\Biggl(\sqrt{1-\sum_m\lambda_m(z)^2/2}+\sum_m\lambda_m(z)\cos[m(\phi-\phi_m)]\Biggr)\, ,
\end{equation} 
where the sums run over the positive integers and the deformation
parameters $\lambda_m(z)$ that represent deviations from axial symmetry are
considered small.

\begin{figure}[!htp]
\begin{center}
\includegraphics[width=0.5\linewidth]{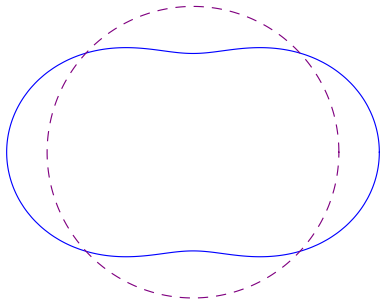}
\end{center}
\caption{Solid line is the deformed cross section at $m=2$}
\label{fig:crosssection}
\end{figure}

Urban {\it et al.\/}~studied deformations with $m\le 6$, but focused mostly
on $m=2$, i.e., quadrupolar deformations.  Deformations of order higher
than $m=6$ cost more surface energy and are therefore less stable. $m=1$
corresponds to a simple translation. One can then consider only deviations
from axisymmetry of the form shown in Fig.~\ref{fig:crosssection}.

\begin{figure}[!htp] 
\begin{center}
\vskip-0.10in
\includegraphics[width=1.0\linewidth]{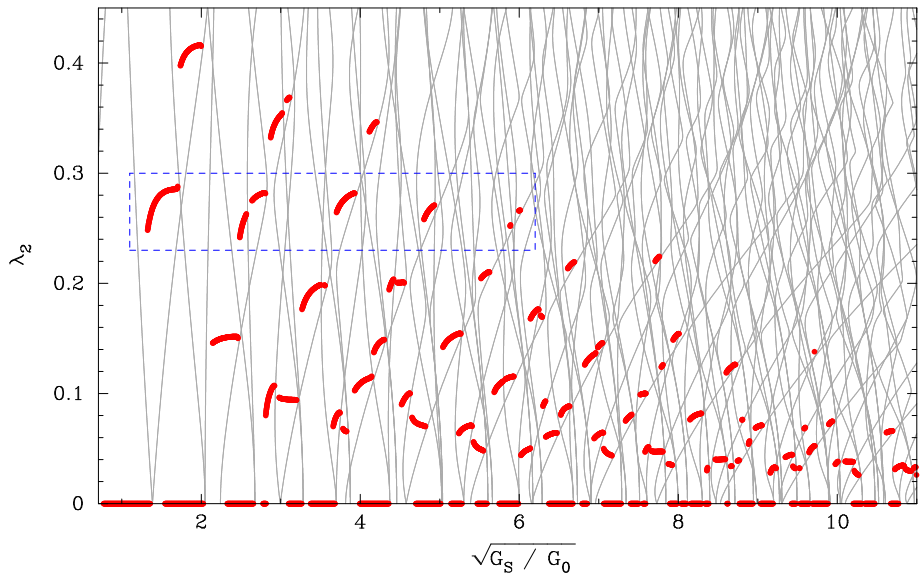}
\end{center}
\vskip-6mm
\caption{Linearly stable quadrupolar Na wires (thick lines) at low
temperature, shown in the deformation configuration
space~$(\rho,\lambda_2)$, where $\rho$ is related to the Sharvin
conductance by $\sqrt{G_s/G_0}\approx k_F\rho/2$.  The thin grey lines show
the thresholds for the openings of new conductance channels.  Thick lines
along the $x$-axis ($\lambda_2=0$) denote regions of stability for purely
cylindrical wires. (From Urban~{\it et al.\/}~\cite{UBSG06}.)}
\label{fig:quadrupolar} 
\end{figure}

Using a linear stability analysis (which again ignores large thermal
fluctuations) they found several sequences of stable wires, some with
considerable deviations from axial symmetry.  Fig.~\ref{fig:quadrupolar}
shows the phase diagram of linear stability in the configuration space of
the two deformation parameters: the Sharvin conductance $G_s$ and the
coefficient of the quadrupole deformation $\lambda_2$.  The $x$-axis gives
a measure of the average radius parameter~$\rho$, which is related to the
square root of the Sharvin~conductance~\cite{SBB97}.

In order to construct a comprehensive theory of nanowire lifetimes, then,
it is necessary to extend the theory developed in~\cite{BSS04, BSS05} to
non-axisymmetric wires.  This requires, as noted earlier, a consideration
of a classical stochastic Ginzburg-Landau field theory with two fields,
with one representing variation of the radius along the longitudinal axis
and the other the departure from axisymmetry.

\section{The Model}
\label{sec:model}

From the preceding discussion, it is evident that a minimal description of
fluctuations in a non-axisymmetric wire requires two fields: the first,
which we denote $\phi_{1}(z)$, is related to $\rho(z)$ and characterizes
radius fluctuations about some fixed average $\rho_0$; the second, which we
denote $\phi_{2}(z)$, is simply $\lambda_{2}(z)$ and characterizes deviations
from axisymmetry. A wire with a perfectly cylindrical cross-section
everywhere would have $\phi_{2}(z)=0$.

Although developing a theory of fluctuations of non-axisymmetric wires
provides a physical motivation for studying classical Ginzburg-Landau
theories with two fields, we are also interested in the general problem of
such field theories. In this paper, we therefore construct and study a
model which applies not only to some (but not all) transitions among
non-axisymmetric wires, but also to other problems as well. This will be
discussed further in Sect.~\ref{sec:discussion}.

\begin{figure}[!htp]
\begin{center}
\includegraphics[width=0.5\linewidth]{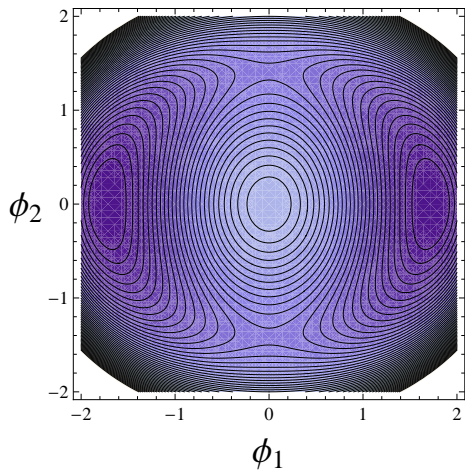}
\end{center}
\caption{Equipotential contours in $U(\phi_1,\phi_2)$ for the fields
$\phi_1$ and $\phi_2$ when $\mu_1=3$ and $\mu_2=2$.}
\label{fig:potential}
\end{figure}

We therefore consider on $[-L/2,L/2]$ two classical fields $\phi_{1}(z,t)$,
$\phi_{2}(z,t)$ subject to the potential
\begin{equation} 
\label{eq:potential}
U(\phi_{1},\phi_{2})=-\frac{\mu_{1}}{2}\phi_{1}^{2}+\frac{1}{4}\phi_{1}^{4}-\frac{\mu_{2}}{2}\phi_{2}^{2}+\frac{1}{4}\phi_{2}^{4}+\frac{1}{2}\phi_{1}^{2}\phi_{2}^{2}
\end{equation}
where $\mu_1$ and $\mu_2$ are arbitrary positive constants such that
$\mu_{1}\neq\mu_{2}$, breaking rotational symmetry between the two fields.
A contour map of the potential is given in Fig.~\ref{fig:potential}.

If the two fields are subject to the potential~(\ref{eq:potential}), have
bending coefficients of $\kappa_1$ and $\kappa_2$ respectively (in the
nanowire case these can be related to surface tension), and are subject to
additive spatiotemporal white noise, then their time evolution is governed
by the pair of coupled stochastic partial differential equations
\begin{eqnarray}
\label{eq:timeevolution}
\dot{\phi_{1}}=\kappa_1\phi_{1}''+\mu_{1}\phi_{1}-\phi_{1}^{3}-\phi_{1}\phi_{2}^{2}+\epsilon^{1/2}\xi_{1}(z,t)\nonumber\\
\dot{\phi_{2}}=\kappa_2\phi_{2}''+\mu_{2}\phi_{2}-\phi_{2}^{3}-\phi_{1}^{2}\phi_{2}+\epsilon^{1/2}\xi_{2}(z,t)\, ,
\end{eqnarray} 
where
$<\xi_{i}(z_{1},t_{1})\xi_{j}(z_{2},t_{2})>=\delta(z_{1}-z_{2})\delta(t_{1}-t_{2})\delta_{ij},~i,j=1,2$.
We will make the simplifying assumption here that
$\kappa_1=\kappa_2=1$. When $\mu_1=\mu_2$, it can be shown that the above
equations are equivalent to those studied by Tarlie {\it et
al.\/}~\cite{Goldbart94} to model fluctuations in superconducting
rings. The breaking of symmetry between the fields leads to an entirely
different behavior from what was found for that case.

The zero-noise dynamics satisfy $\dot{\phi_{i}}=-\delta H/\delta\phi_{i}$,
with the energy functional 
\begin{equation}
H=\int_{-L/2}^{L/2} \!
(\frac{1}{2}(\phi_{1}'(z))^{2}+\frac{1}{2}(\phi_{2}'(z))^{2}+U(\phi_{1},\phi_{2}))
\, dz\, . 
\label{eq:H}
\end{equation}
The metastable and saddle, or transition, states are time-independent
solutions of the zero-noise equations,~\cite{Hanggi90} satisfying the
Euler-Lagrange equations
\begin{eqnarray}
\label{eq:EL}
\phi_{1}''=-\mu_{1}\phi_{1}+\phi_{1}^{3}+\phi_{1}\phi_{2}^{2}\nonumber\\
\phi_{2}''=-\mu_{2}\phi_{2}+\phi_{2}^{3}+\phi_{1}^{2}\phi_{2} 
\end{eqnarray}

At nonzero temperature, thermal fluctuations can drive the system from one
metastable state to another. Such a transition proceeds via a pathway of
states that first goes uphill in energy from the starting configuration,
passes through (or close to) a saddle configuration, and then proceeds
downhill towards the nearest metastable state. The activation rate is given
in the $T\to0$ limit by the Kramers formula~\cite{Hanggi90}
\begin{equation}
\label{eq:Kramers}
\Gamma\sim \Gamma_0\exp(-\Delta E/T)\, .
\end{equation}
Here $\Delta E$ is the activation barrier, the difference in energy between
the saddle and the starting metastable configuration, and $\Gamma_0$ is the
rate prefactor.

The quantities $\Delta E$ and $\Gamma_0$ depend on the details of the
potential~(\ref{eq:potential}), on the interval length~$L$ on which the
fields are defined, and on the choice of boundary conditions at the
endpoints $z=-L/2$ and~$z=L/2$.  It was shown in~\cite{Burki03} that Neumann
boundary conditions are appropriate for the nanowire problem, and we will
use them here.  However, the theory is easily extended to other types of
boundary conditions~\cite{MSspie}, although for periodic boundary
conditions care must be taken to extract the zero mode when performing
prefactor computations~\cite{McKane95}.

From here on, we choose without loss in generality $\mu_1>\mu_2$. In this
case there are two uniform metastable states
$\phi_{1s}=\pm\sqrt{\mu_{1}},\phi_{2s}=0$ and two uniform transition states
$\phi_{1u}=0,\phi_{2u}=\pm\sqrt{\mu_{2}}$, as can be seen in
Fig.~\ref{fig:potential}. In the following section, we will see that the
uniform transition states are true saddles, and therefore relevant for
escape, only when $L<L_c=\frac{\pi}{\sqrt{\mu_{1}-\mu_{2}}}$. At $L_c$, a
transition occurs, and above it the transition states are nonuniform.

\section{The Transition State}
\label{sec:transition}

\begin{figure}[!htp]
\centering
\subfloat[$\phi^{\rm inst}_{1,m}(z)$]{\label{fig:ins1}\includegraphics[width=0.4\textwidth]{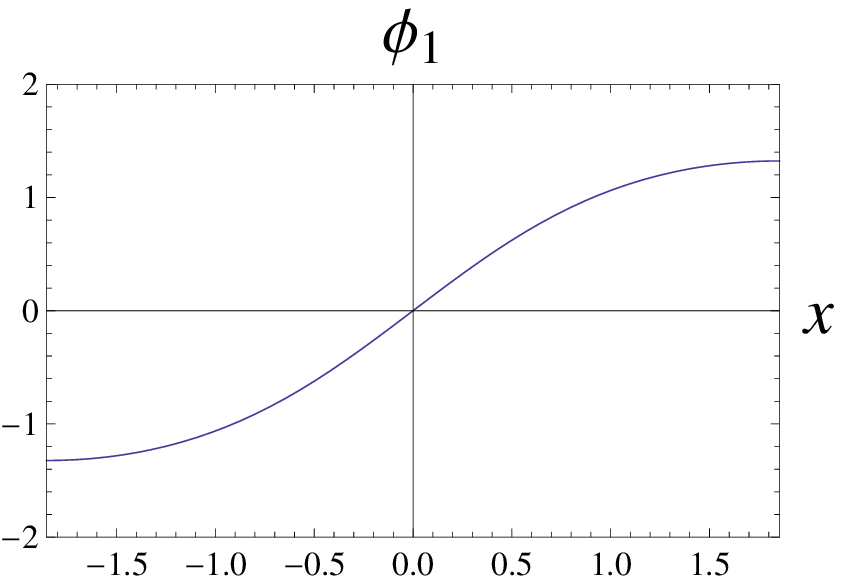}}
\subfloat[$\phi^{\rm inst}_{2,m}(z)$]{\label{fig:ins2}\includegraphics[width=0.4\textwidth]{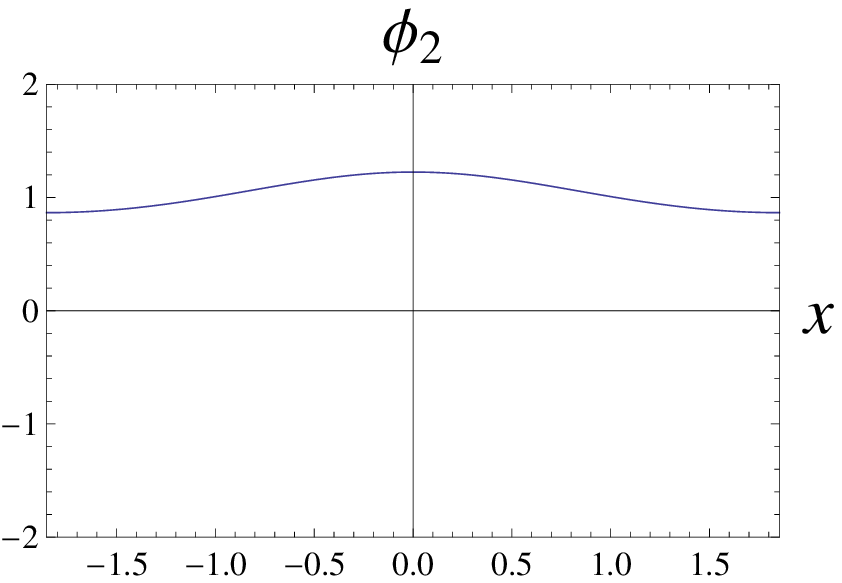}}
\caption{The instanton states $\phi^{\rm inst}_{i,m}$ for $\mu_{1}=3$,
    $\mu_{2}=2$, and $m=1/2$.}
\label{fig:instantons}
\end{figure}

We have found analytical solutions to Eqs.~(\ref{eq:EL}) that describe
nonuniform saddle configurations (hereafter referred to as ``instantons'',
in keeping with the usual practice).  For general $\mu_1>\mu_2$ they are:
\begin{gather}
\phi^{\rm
inst}_{1,m}(z)=\pm\sqrt{m}\sqrt{(2\mu_{1}-\mu_{2})-m(\mu_{1}-\mu_{2})}{\rm sn}(\sqrt{\mu_{1}-\mu_{2}}\,z|m)
\label{eq:ins1}\\ \phi^{\rm
inst}_{2,m}(z)=\pm\sqrt{\mu_{2}-m(\mu_{1}-\mu_{2})}{\rm dn}(\sqrt{\mu_{1}-\mu_{2}}\,z|m)
\label{eq:ins2}
\end{gather}
where ${\rm sn}(.|m)$and ${\rm dn}(.|m)$ are the Jacobi elliptic functions with
parameter $m$, whose periods are $4K(m)$ and $2K(m)$ respectively, with
$K(m)$ the complete elliptic integral of the first
kind~\cite{Abramowitz65}. Imposing Neumann boundary conditions yields a
relation between $L$ and $m$:
\begin{equation} \label{eq:lc}
L=\frac{2K(m)}{\sqrt{\mu_{1}-\mu_{2}}}
\end{equation}
The instanton states for $\mu_1=3$, $\mu_2=2$ and intermediate $m$ are
shown in~Fig.~\ref{fig:instantons}.  

As $m\to 0$, $L$ approaches its minimum length
$L_{c}=\frac{\pi}{\sqrt{\mu_{1}-\mu_{2}}}$. In this limit, $\phi^{\rm
inst}_{1,0}=0$, $\phi^{\rm inst}_{2,0}=\pm\sqrt{\mu_{2}}$, and the
instantons reduce to the spatially uniform saddle states. As $m \to 1^{-}$,
$L \to \infty$, and the instanton states become
\begin{gather}
\phi^{\rm inst}_{1,1}=\pm\sqrt{\mu_{1}}\, \tanh(\sqrt{\mu_{1}-\mu_{2}}\, z)
\label{eq:tanh} \\ \phi^{\rm
inst}_{2,1}=\pm\sqrt{2\mu_{2}-\mu_{1}}\, {\rm sech}(\sqrt{\mu_{1}-\mu_{2}} \, z)
\label{eq:sech}
\end{gather}
\begin{figure}[!htp]
\centering
\subfloat[$\phi^{\rm inst}_{1,m}(z)$]{\label{fig:tanh}\includegraphics[width=0.4\textwidth]{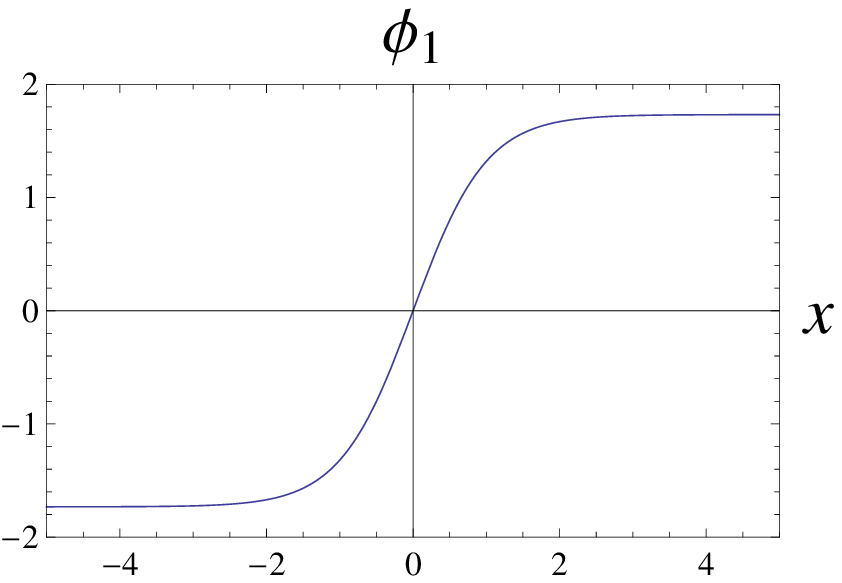}}
\subfloat[$\phi^{\rm inst}_{2,m}(z)$]{\label{fig:sech}\includegraphics[width=0.4\textwidth]{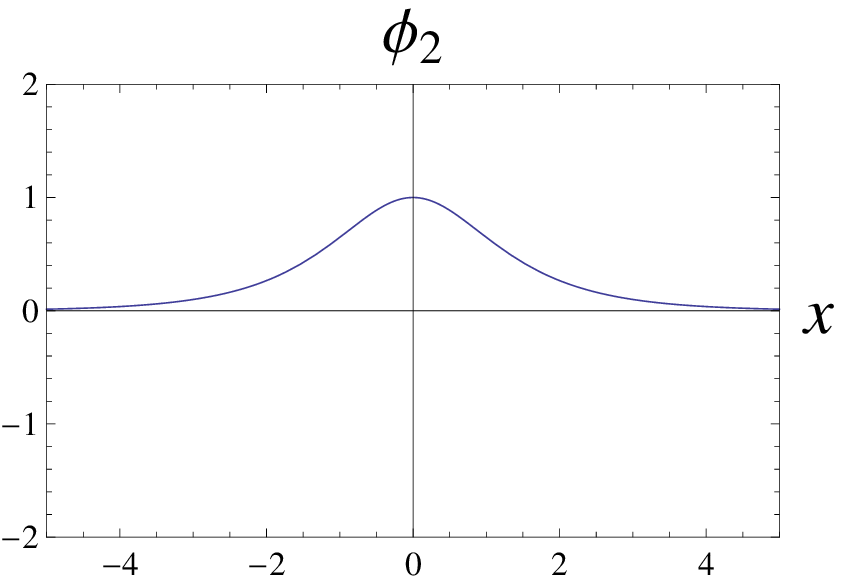}}
\caption{The instanton states $\phi^{\rm inst}_{i,m}$ for infinite length as $m\to1^{-}$ for $\mu_{1}=3$, $\mu_{2}=2$.}
\label{fig:tanhsech}
\end{figure}
In the nanowire case, there is a nice geometric interpretation of this
particular version of the escape process, which will be discussed in
Sect.~\ref{sec:discussion}.

\begin{figure}[!htp]
\centering
\includegraphics[width=\linewidth]{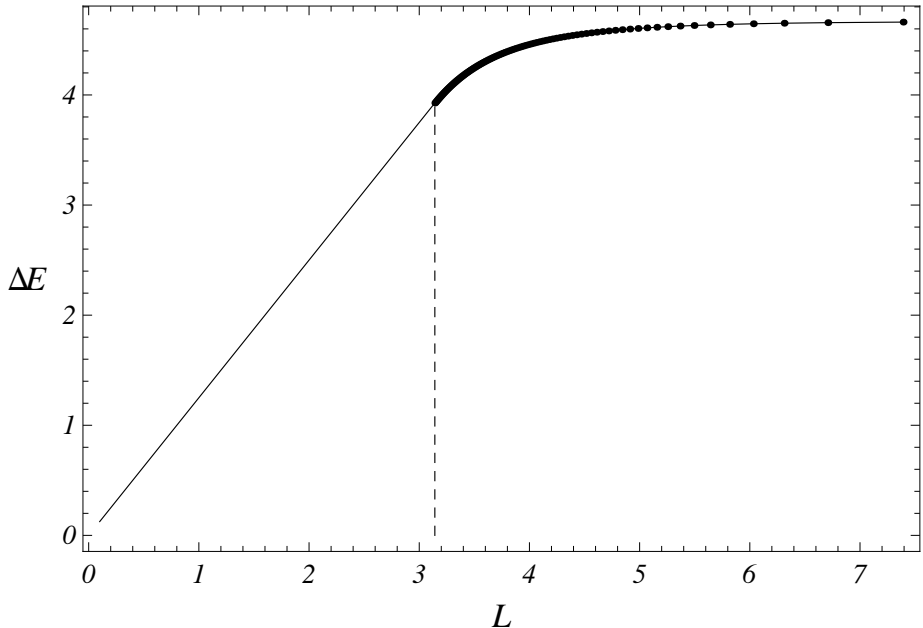}
\caption{The activation barrier $\Delta E$ for Neumann boundary condition (solid line) at $\mu_1=3,\mu_2=2$. The dashed line indicates the crossover from the uniform transition states to the instanton transition states at $L_{c}$. The dots represent the numerical results of $\Delta E$ beyond $L_{c}$.}
\label{fig:barrier}
\end{figure}

At low temperatures, the leading order asymptotic dependence ($\Delta E$ in
the Kramers formula) of the escape rate can be computed as the difference
in energy between the saddle and stable configurations, and is plotted in
Fig.~\ref{fig:barrier}.\\
As $L \to \infty$, $\Delta E$ approaches
$\frac{2}{3}\sqrt{\mu_{1}-\mu_{2}}(\mu_{1}+2\mu_{2})$.  This is simply the
energy of the domain walls of Figs.~\ref{fig:tanh},~\ref{fig:sech}.

\section{Rate Prefactor}
\label{sec:prefactor}

For an overdamped system driven by white noise, the rate prefactor
$\Gamma_{0}$ can be derived in principle~\cite{Hanggi90}, although this is
often difficult in practice. The usual procedure is to consider small
perturbations $\eta_{1}$, $\eta_{2}$ about the metastable state:
$\phi_{1}=\phi_{1,s}+\eta_{1}$ and $\phi_{2}=\phi_{2,s}+\eta_{2}$. Then to
leading order $\dot{\vec{\eta}}=-{\bf\Lambda}_{s}{\vec{\eta}}$,
${\vec{\eta}}=(\eta_{1},\eta_{2})$, where ${\bf\Lambda}_{s}$ is the linearized
zero-noise dynamical operator at $\phi_{1,s}$,
$\phi_{2,s}$. Similarly, ${\bf\Lambda}_{u}$ is the linearized zero-noise
dynamical operator at the transition state
$\phi_{1,u}$,$\phi_{2,u}$. Then~\cite{Hanggi90}
\begin{equation}
\label{eq:prefactor}
\Gamma_{0}=\frac{1}{\pi}\sqrt{\left|\frac{\det\,{\bf\Lambda}_{s}}{\det\,{\bf\Lambda}_{u}}\right|}|\lambda_{u,1}|
\end{equation}
where $\lambda_{u,1}$ is the single negative eigenvalue of
${\bf\Lambda}_{u}$, corresponding to the direction along which the optimal
escape trajectory approaches the transition state.  Here
Eq.~(\ref{eq:prefactor}) differs from the usual formula for $\Gamma_0$~\cite{Gardiner85,Hanggi90} by
a factor of 2 in the denominator because we have two saddle points
distributed symmetrically between the metastable states (see
Fig.~\ref{fig:potential}) such that the transition can take place via
either. In the next two sections, we consider two interval length regimes,
each with different saddle configurations.

\subsection{$L<L_{c}$}
\label{subsec:below}

In this regime, both the metastable and transition states are spatially
uniform, allowing for a straightforward computation of $\Gamma_{0}$. Assume
the system begins at the metastable state $(-\sqrt{\mu_{1}},0)$, passes
through the transition state $(0,\sqrt{\mu_{2}})$, and finishes at
$(\sqrt{\mu_{1}},0)$. Linearization around the metastable state gives
\begin{equation} \label{eq:lambda1}
\frac{d}{dt}
\begin{pmatrix}
\eta_{1}\\
\eta_{2}
\end{pmatrix}
=-{\bf\Lambda}_{s}
\begin{pmatrix}
\eta_{1}\\
\eta_{2}
\end{pmatrix}
=-
\begin{pmatrix}
-\partial_{z}^{2}+2\mu_{1} & 0\\
0 & -\partial_{z}^{2}+(\mu_{1}-\mu_{2})
\end{pmatrix}
\begin{pmatrix}
\eta_{1}\\
\eta_{2}
\end{pmatrix}
\end{equation}
and around the transition state
\begin{equation} \label{eq:lambda2}
\frac{d}{dt}
\begin{pmatrix}
\eta_{1}\\
\eta_{2}
\end{pmatrix}
=-{\bf\Lambda}_{u}
\begin{pmatrix}
\eta_{1}\\
\eta_{2}
\end{pmatrix}
=-
\begin{pmatrix}
-\partial_{z}^{2}-(\mu_{1}-\mu_{2}) & 0\\
0 & -\partial_{z}^{2}+2\mu_{2}
\end{pmatrix}
\begin{pmatrix}
\eta_{1}\\
\eta_{2}
\end{pmatrix}
\, .
\end{equation}

The spectrum corresponding to ${\bf\Lambda}_{s}$ is
\[ \label{eq:spectrum1}
\lambda_{n}^{s}=\left\{
\begin{array}{l l}
\frac{\pi^{2}n_{1}^{2}}{L^{2}}+2\mu_{1} & \quad  n_{1}=0,1,2...\\
\\
\frac{\pi^{2}n_{2}^{2}}{L^{2}}+(\mu_{1}-\mu_{2}) & \quad n_{2}=0,1,2...\\
\end{array} \right.
\]
and that corresponding to ${\bf\Lambda}_{u}$ is
\[ \label{eq:spectrum2}
\lambda_{n}^{u}=\left\{
\begin{array}{l l}
\frac{\pi^{2}n_{1}^{2}}{L^{2}}-(\mu_{1}-\mu_{2}) & \quad n_{1}=0,1,2...\\
\\
\frac{\pi^{2}n_{2}^{2}}{L^{2}}+2\mu_{2} & \quad n_{2}=0,1,2...\\
\end{array} \right.
\]

When $L<L_c$, all the eigenvalues of ${\bf\Lambda}_{u}$ are positive,
except for $\lambda_{n_{1}=0}^{u}=-(\mu_{1}-\mu_{2})$, indicating that this
is indeed a saddle configuration. The eigenfunction corresponding to
$\lambda_{n_{1}=0}^{u}$, which is spatially uniform, is the direction in
configuration space along which the optimal escape path approaches
$(0,\sqrt{\mu_2})$. The fact that the lowest positive eigenvalue
$\lambda_{n_{1}=1}^{u}\to 0$ as $L\to L_c^-$ indicates a transition in the
escape dynamics at $L_c$.

This in turn affects the rate prefactor, which for $L<L_{c}$ is
\begin{gather} \label{eq:prefactor1}
\begin{split}
\Gamma_{0}=&\frac{1}{\pi}\sqrt{\left|\frac{\prod_{n_1=0}^{\infty}\left(2\mu_{1}+\frac{\pi^{2}n_1^{2}}{L^{2}}\right)}{\prod_{n_2=0}^{\infty}\left(2\mu_{2}+\frac{\pi^{2}n_2^{2}}{L^{2}}\right)}\right|}\sqrt{\left|\frac{\prod_{n_2=0}^{\infty}\left((\mu_{1}-\mu_{2})+\frac{\pi^{2}n_2^{2}}{L^{2}}\right)}{\prod_{n_1=0}^{\infty}\left(-(\mu_{1}-\mu_{2})+\frac{\pi^{2}n_1^{2}}{L^{2}}\right)}\right|}|-(\mu_{1}-\mu_{2})|\\
=&\frac{1}{\pi}\sqrt{\frac{\sqrt{\mu_{1}}}{\sqrt{\mu_{2}}}}\frac{\sqrt{\sinh(\sqrt{\mu_{1}-\mu_{2}}L)}}{\sqrt{\sinh(\sqrt{2\mu_{2}}L)}}\frac{\sqrt{\sinh(\sqrt{2\mu_{1}}L)}}{\sqrt{|\sin(\sqrt{\mu_{1}-\mu_{2}}L)|}}(\mu_{1}-\mu_{2})\, .
\end{split}
\end{gather}

The prefactor diverges at $L_{c}=\frac{\pi}{\sqrt{\mu_{1}-\mu_{2}}}$. The
divergence arises from the vanishing of
$\lambda_{n_{1}=1}^{u}=\frac{\pi^{2}}{L^{2}}-(\mu_{1}-\mu_{2})$, indicating
an appearance of a transverse soft mode in the fluctuations about the
transition state.

\subsection{$L>L_{c}$}
\label{subsec:above}

When the transition is nonuniform, computation of the determinant ratio is
less straightforward.  Here the linearized operator ${\bf\Lambda}_{u}$ about the
transition state is
\begin{equation} \label{eq:lambda3}
{\bf\Lambda}_{u}=
\begin{pmatrix}
-\partial_{z}^{2}-(\mu_{1}-3(\phi^{\rm inst}_{1,m})^{2}-(\phi^{\rm inst}_{2,m})^{2}) & 2\phi^{\rm inst}_{1,m}\phi^{\rm inst}_{2,m}\\
2\phi^{\rm inst}_{1,m}\phi^{\rm inst}_{2,m} & -\partial_{z}^{2}-(\mu_{2}-3(\phi^{\rm inst}_{2,m})^{2}-(\phi^{\rm inst}_{1,m})^{2})
\end{pmatrix}
\end{equation}
and the linearized equations become a pair of second order coupled
nonlinear differential equations. 

In order to calculate $\det{\bf{\bf\Lambda}}_{s}/\det{\bf{\bf\Lambda}}_{u}$
we make use of a generalization, due to Forman~\cite{Forman87}, of the
Gel'fand-Yaglom technique~\cite{gelfand60}, suitable for differential
operators in a $2 \times 2$-matrix. The Forman method is readily
extendible to higher dimensions~\cite{mckane03}, and its central result is
that
\begin{equation} \label{eq:quotient}
\frac{\det{\bf\Lambda}_{s}}{\det{\bf\Lambda}_{u}}=\frac{\det[{\bf M}+{\bf N
      Y}_{s}(L/2)]}{\det[{\bf M}+{\bf N Y}_{u}(L/2)]}\, .
\end{equation}
The $4 \times 4$ matrices ${\bf Y}_{s}(z)$, ${\bf Y}_{u}(z)$
in~(\ref{eq:quotient}) are ``fundamental'' matrices\cite{hartman82}.  A
fundamental matrix ${\bf Y}(z)$ has the property that, for any solution
${\vec{\eta}}(z)$ of the homogeneous equation ${\bf\Lambda}{\vec{\eta}}=0$,
\begin{equation} \label{eq:defineY}
\begin{pmatrix}
\eta_{1}(z)\\
\eta_{2}(z)\\
\eta_{1}'(z)\\
\eta_{2}'(z)
\end{pmatrix}
={\bf Y}(z)
\begin{pmatrix}
\eta_{1}(-L/2)\\
\eta_{2}(-L/2)\\
\eta_{1}'(-L/2)\\
\eta_{2}'(-L/2)\, .
\end{pmatrix}
\end{equation} 

The matrices ${\bf Y}_{s}(z)$, ${\bf Y}_{u}(z)$ are then defined to be fundamental matrices of the
differential equations
\begin{equation} \label{eq:ode1}
\frac{d}{dz}
\begin{pmatrix}
\eta_{1}\\
\eta_{2}\\
\eta_{1}'\\
\eta_{2}'
\end{pmatrix}
=
\begin{pmatrix}
0 & 0 & 1 & 0\\
0 & 0 & 0 & 1\\
2\mu_{1} & 0 & 0 & 0\\
0 & \mu_{1}-\mu_{2} & 0 & 0
\end{pmatrix}
\begin{pmatrix}
\eta_{1}\\
\eta_{2}\\
\eta_{1}'\\
\eta_{2}'
\end{pmatrix}
\end{equation}
\begin{equation} \label{eq:ode2}
\frac{d}{dz}
\begin{pmatrix}
\eta_{1}\\
\eta_{2}\\
\eta_{1}'\\
\eta_{2}'
\end{pmatrix}
=
\begin{pmatrix}
0 & 0 & 1 & 0\\
0 & 0 & 0 & 1\\
-(\mu_{1}-3(\phi^{\rm inst}_{1,m})^{2}-(\phi^{\rm inst}_{2,m})^{2}) & 2\phi^{\rm inst}_{1,m}\phi^{\rm inst}_{2,m} & 0 & 0\\
2\phi^{\rm inst}_{1,m}\phi^{\rm inst}_{2,m} & -(\mu_{2}-3(\phi^{\rm inst}_{2,m})^{2}-(\phi^{\rm inst}_{1,m})^{2}) & 0 & 0
\end{pmatrix}
\begin{pmatrix}
\eta_{1}\\
\eta_{2}\\
\eta_{1}'\\
\eta_{2}'
\end{pmatrix}
\end{equation}
which are just the equivalent first order versions of ${\bf\Lambda}_{s}$
and ${\bf\Lambda}_{u}$ (for $L>L_{c}$).  The matrices ${\bf Y}_{s}(z)$,
${\bf Y}_{u}(z)$ will be discussed further below.

The $4 \times 4$ matrices ${\bf M}$ and ${\bf N}$ encode the boundary
conditions
\begin{gather}
{\bf M}\,
\begin{pmatrix}
{\vec{\eta}}(-L/2)\\
{\vec{\eta}}\,'(-L/2)
\end{pmatrix}
+{\bf N}\,
\begin{pmatrix}
{\vec{\eta}}(L/2)\\
{\vec{\eta}}\,'(L/2)
\end{pmatrix}
=0\, .
\label{eq:MN1}
\intertext{For Neumann boundary conditions, ${\bf M}$ and ${\bf N}$ have the forms}
{\bf M}=\begin{pmatrix}
0 & 0 & 1 & 0\\
0 & 0 & 0 & 1\\
0 & 0 & 0 & 0\\
0 & 0 & 0 & 0
\end{pmatrix}
\intertext{and}
%\begin{pmatrix}
%\eta_{1}(-L/2)\\
%\eta_{2}(-L/2)\\
%\eta_{1}'(-L/2)\\
%\eta_{2}'(-L/2)
%\end{pmatrix}
%+
{\bf N}=\begin{pmatrix}
0 & 0 & 0 & 0\\
0 & 0 & 0 & 0\\
0 & 0 & 1 & 0\\
0 & 0 & 0 & 1
\end{pmatrix}
%\begin{pmatrix}
%\eta_{1}(L/2)\\
%\eta_{2}(L/2)\\
%\eta_{1}'(L/2)\\
%\eta_{2}'(L/2)
%\end{pmatrix}
%=
%\begin{pmatrix}
%0\\
%0\\
%0\\
%0
%\end{pmatrix} 
\label{eq:MN2}
\end{gather}
Thus, the determinant ratio of two infinite-dimensional matrices is reduced
to the ratio of the determinants of two finite-dimensional matrices.\\ 

The most difficult part of the strategy outlined in the previous paragraph
is computation of the fundamental matrix.  We proceed as follows.  The
matrix ${\bf Y}(z)$ can be expressed as the product of ${\bf H}(z){\bf
H}^{-1}(-L/2)$, where ${\bf H}(z)$ is constructed from the four linearly
independent solutions of the first order equation~(\ref{eq:ode1}) for the
metastable state, or~(\ref{eq:ode2}) for the transition state. Suppose\\
\begin{center}
$
\begin{pmatrix}
h_{1}(z)\\
h_{2}(z)\\
h_{1}'(z)\\
h_{2}'(z)
\end{pmatrix}
$
,
$
\begin{pmatrix}
h_{3}(z)\\
h_{4}(z)\\
h_{3}'(z)\\
h_{4}'(z)
\end{pmatrix}
$
,
$
\begin{pmatrix}
h_{5}(z)\\
h_{6}(z)\\
h_{5}'(z)\\
h_{6}'(z)
\end{pmatrix}
$
,
$
\begin{pmatrix}
h_{7}(z)\\
h_{8}(z)\\
h_{7}'(z)\\
h_{8}'(z)
\end{pmatrix}
$
\end{center}
are the four independent solutions, each satisfying (using the transition
state for specificity)
\begin{equation}
\frac{d}{dz}
\begin{pmatrix}
h_{1}\\
h_{2}\\
h_{1}'\\
h_{2}'
\end{pmatrix}
=
\begin{pmatrix}
0 & 0 & 1 & 0\\
0 & 0 & 0 & 1\\
-(\mu_{1}-3(\phi^{\rm inst}_{1,m})^{2}-(\phi^{\rm inst}_{2,m})^{2}) & 2\phi^{\rm inst}_{1,m}\phi^{\rm inst}_{2,m} & 0 & 0\\
2\phi^{\rm inst}_{1,m}\phi^{\rm inst}_{2,m} & -(\mu_{2}-3(\phi^{\rm inst}_{2,m})^{2}-(\phi^{\rm inst}_{1,m})^{2}) & 0 & 0
\end{pmatrix}
\begin{pmatrix}
h_{1}\\
h_{2}\\
h_{1}'\\
h_{2}'
\end{pmatrix}
\end{equation}
or equivalently
\begin{equation}
{\bf\Lambda}_{u}\,
\begin{pmatrix}
h_{1}(z)\\
h_{2}(z)
\end{pmatrix}
=
\begin{pmatrix}
-\partial_{z}^{2}-(\mu_{1}-3(\phi^{\rm inst}_{1,m})^{2}-(\phi^{\rm inst}_{2,m})^{2}) & 2\phi^{\rm inst}_{1,m}\phi^{\rm inst}_{2,m}\\
2\phi^{\rm inst}_{1,m}\phi^{\rm inst}_{2,m} & -\partial_{z}^{2}-(\mu_{2}-3(\phi^{\rm inst}_{2,m})^{2}-(\phi^{\rm inst}_{1,m})^{2})
\end{pmatrix}
\begin{pmatrix}
h_{1}(z)\\
h_{2}(z)
\end{pmatrix}
=0
\end{equation}
and similarly for the rest. Then ${\bf H}(z)$ is:
\begin{equation}
{\bf H}(z)=
\begin{pmatrix}
h_{1}(z) & h_{3}(z) & h_{5}(z) & h_{7}(z)\\
h_{2}(z) & h_{4}(z) & h_{6}(z) & h_{8}(z)\\
h_{1}'(z)& h_{3}'(z)& h_{5}'(z)& h_{7}'(z)\\
h_{2}'(z)& h_{4}'(z)& h_{6}'(z)& h_{8}'(z)\, . 
\end{pmatrix}
\end{equation}

While it's elementary to obtain the independent solutions of
${\bf\Lambda}_{s}{\vec{\eta}}_{s}=0$ at the metastable state, there's no
systematic way of dealing with nonuniform transition states.  In our
problem, we note (cf.~(\ref{eq:quotient})) that it is sufficient to compute
the fundamental matrix ${\bf Y}(z)$ at the boundary $z= L/2$.  One can
therefore numerically integrate the coupled differential equations forward
from $z=-L/2$ using~(\ref{eq:ode2}) with four independent initial values,
as follows.

Suppose that $(\eta_{1}(z),\eta_{2}(z),\eta_{1}'(z),\eta_{2}'(z))$ is any
solution satisfying~(\ref{eq:ode2}), $\alpha,\beta,\gamma,\delta$ are four
arbitrary constants, and ${\bf H}(z)$ is the matrix of the four linearly
independent solution vectors. Then it follows that
\begin{equation} \label{eq:generalsolu}
\begin{pmatrix}
\eta_{1}(z)\\
\eta_{2}(z)\\
\eta_{1}'(z)\\
\eta_{2}'(z)
\end{pmatrix}
={\bf H}(z)
\begin{pmatrix}
\alpha\\
\beta\\
\gamma\\
\delta
\end{pmatrix}\, .
\end{equation}
Inverting, we get
\begin{equation} \label{eq:invertedsolu}
\begin{pmatrix}
\alpha\\
\beta\\
\gamma\\
\delta
\end{pmatrix}
={\bf H}^{-1}(z)
\begin{pmatrix}
\eta_{1}(z)\\
\eta_{2}(z)\\
\eta_{1}'(z)\\
\eta_{2}'(z)
\end{pmatrix}\, .
\end{equation}
Since $\alpha,\beta,\gamma,\delta$ are arbitrary constants, we can write
without loss of generality
\begin{equation} \label{eq:invertedsolu2}
\begin{pmatrix}
\alpha\\
\beta\\
\gamma\\
\delta
\end{pmatrix}
={\bf H}^{-1}(-L/2)
\begin{pmatrix}
\eta_{1}(-L/2)\\
\eta_{2}(-L/2)\\
\eta_{1}'(-L/2)\\
\eta_{2}'(-L/2)
\end{pmatrix}\, .
\end{equation}
Using this to replace the constants in Eq.~(\ref{eq:generalsolu}), we arrive at
\begin{equation} \label{eq:ymatrix}
\begin{pmatrix}
\eta_{1}(z)\\
\eta_{2}(z)\\
\eta_{1}'(z)\\
\eta_{2}'(z)
\end{pmatrix}
={\bf H}(z)\,{\bf H}^{-1}(-L/2)
\begin{pmatrix}
\eta_{1}(-L/2)\\
\eta_{2}(-L/2)\\
\eta_{1}'(-L/2)\\
\eta_{2}'(-L/2)
\end{pmatrix}
\end{equation}
which is simply~(\ref{eq:defineY}).  It is now clear that only the boundary
values of ${\bf H}(z)$ at $z=\pm L/2$ are needed to compute ${\bf Y}(L/2)$. For example, we
can choose
\begin{equation} \label{eq:identitymatrix}
\begin{pmatrix}
1 & 0 & 0 & 0\\
0 & 1 & 0 & 0\\
0 & 0 & 1 & 0\\
0 & 0 & 0 & 1
\end{pmatrix}
\end{equation}
for ${\bf H}(-L/2)$, and numerically integrate forward using~(\ref{eq:ode2})
to get ${\bf H}(L/2)$. Since each column of the identity matrix~(\ref{eq:identitymatrix}) is
independent from the others, the columns in ${\bf H}(L/2)$ are also
linearly independent. In this way, ${\bf Y}(L/2)$ is readily obtained.

\begin{figure}[!htp]
\begin{center}
\includegraphics[width=\linewidth]{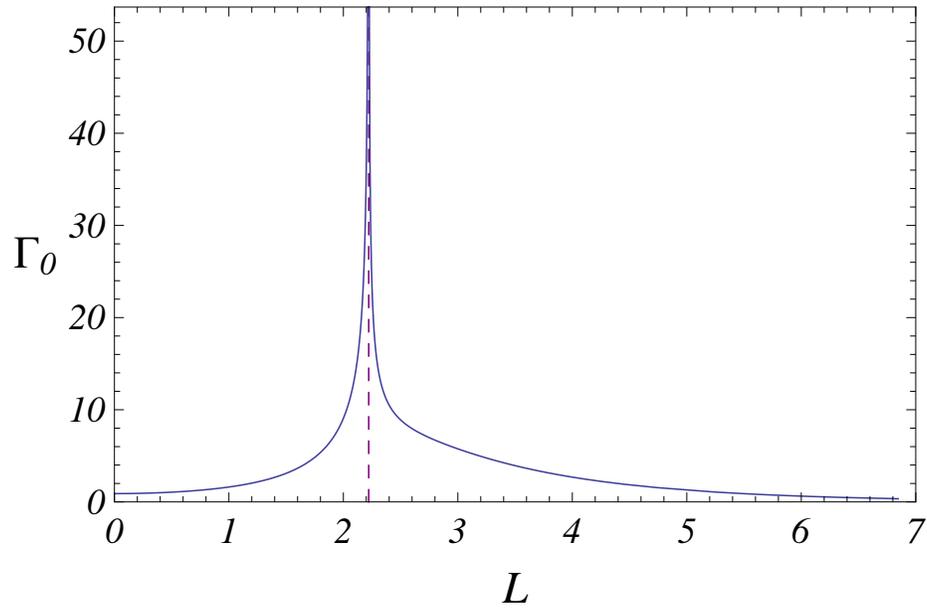}
\end{center}
\caption{Prefactor $\Gamma_0$ vs.~$L$ for $\mu_{1}=4$, $\mu_{2}=2$.}
\label{fig:prefactor}
\end{figure}

\begin{figure}[!htp]
\begin{center}
\includegraphics[width=\linewidth]{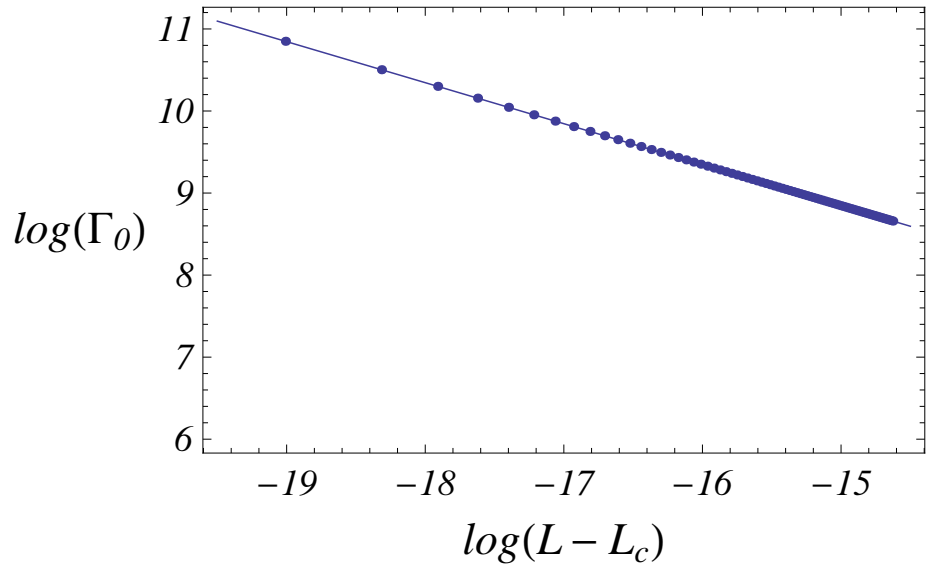}
\end{center}
\caption{$\log\Gamma_0$ vs.~$\log (L-L_{c})$ near $L_c^+$ for
$\mu_{1}=4$, $\mu_{2}=2$. The dots indicate numerical data. The line fit
gives $\log\Gamma_0=-0.5\log (L-L_{c})+1.35$ for $L$ very close to (and above)
$L_c$.}
\label{fig:cexpo}
\end{figure}

Fig.~\ref{fig:prefactor} shows the prefactor $\Gamma_{0}$ vs.~$L$. The
divergence of the prefactor as $L\to L_c$ from either side is preserved in
the two-field case; its physical meaning has been discussed
in~\cite{Stein04}. The critical exponent characterizing the divergence is
of interest and can be readily computed. When $L<L_{c}$, it is easy to see
from~(\ref{eq:prefactor1}) that
\begin{equation} \label{eq:cexpo1}
\Gamma_{0}\sim (L-L_{c})^{-\frac{1}{2}}\, .
\end{equation}
For $L>L_c$, the critical exponent can be computed numerically and is also
$1/2$, as shown in Fig.~\ref{fig:cexpo}.


\section{Discussion}
\label{sec:discussion}

We have introduced a stochastic Ginzburg-Landau model with two coupled
fields, and have found explicit solutions for the nonuniform saddle states
that govern noise-induced transitions from one stable state to another.
This model has a phase transition as system size changes, similar to what
has been seen in the single-field case~\cite{MS01b,Stein04,MS03,Stein05}.

The action functional~(\ref{eq:H}) was designed to model nanowire stability
and decay when non-axisymmetric cross sections appear during the time
evolution of the wire, either as beginning, ending, or intermediate
states. However, the model --- or its modifications --- has wider
applicability, which we briefly discuss below.  We note also that the model
can be viewed as a generalization of that used by Tarlie {\it et
al.\/}~\cite{Goldbart94} to model transitions between different conducting
states in superconducting rings.

As noted in Sect.~\ref{sec:intro}, the system analyzed here can serve to
model certain transitions among different nanowire states, but not all.
Nevertheless, it provides a nice physical interpretation of the escape
process in non-axisymmetric nanowires. The escape process occurs, as usual,
via nucleation of a ``droplet'' of one metastable configuration in the
background of the other. Fig.~\ref{fig:tanhsech} indicates that the
infinite nanowire begins as a cylindrically symmetric wire at one (uniform)
radius and ends as a cylindrically symmetric wire at a different radius,
while passing through a sequence of nonaxisymmetric configurations.  It is
easy to see that if we were to take instead $\mu_2>\mu_1$, the model would
describe a process where the starting and ending states have the same
average radius but are non-axisymmetric with different deformation states,
while the saddle is cylindrically symmetric.  By varying the model
parameters, one can interpolate between these two extreme cases.  An
extensive analysis of such transitions, including applications of the
current model, will appear in~\cite{BGSSinprep}.

The model presented here, or ones close to it, apply also to a variety of
other situations. One of these is constrained dynamics of the sort that
plays a role in viscous, or glassy, liquids (see, for
example,~\cite{PSAA84, FA84}).  The simplest situation corresponding to
this sort of dynamics is described in Munoz~{\it et al.\/}~\cite{MGIP98}.
Fig.~2 of that paper shows a particle attempting to diffuse from one stable
position to another; if a second particle is in one of {\it its\/} two
stable positions, it blocks the first.  Aside from its natural occurrence
in glassy liquids, such a situation may also be created and studied using
an optical trap with metastable wells, such as that employed by McCann~{\it
et al.\/}~\cite{McCann99} or Seol~{\it et al.\/}~\cite{SSV09} to test the
Kramers transition rate formula under varying conditions.

If one removes the diffusion terms from Eqs.~(\ref{eq:timeevolution}),
treating $\phi_1$ and $\phi_2$ simply as spatial coordinates, then one has
a model where two degrees of freedom act on each other in a similar manner,
although in a mutual fashion (see, for example,~\cite{MS92}). However,
inclusion of the diffusion terms describes a more interesting situation.
Here one can think of the field $\phi_1(z)$ as describing a density of a
certain particle species along an interval, and $\phi_2(z)$ as representing
a second species density.  Diffusion of particles into or out of the
interval depends on the state of $\phi_2(z)$; its vanishing (with the set
of parameters used in this paper) ``locks'' the density of $\phi_1(z)$ to
one of its stable configurations.  In order for any diffusion to take
place, $\phi_2(z)$ must change also, but its density distribution is
coupled to that of $\phi_1(z)$.  It would be of interest to pursue this
further to investigate spatially {\it inhomogeneous\/} models of
constrained dynamics.

There are a number of other physical situations that can be stochastically
modelled using two coupled classical fields. Examples of such models are
the Fitzhugh-Nagumo model~\cite{Fitzhugh55,NAY62} describing excitable
media in general (and neurons in particular), and the Gray-Scott
model~\cite{Pearson93} of chemical reaction-diffusion systems.  These,
however, are nonpotential systems, and can exhibit phenomena, such as limit
cycles, that potential systems cannot.  Nevertheless, if one examines
transitions between ``fixed points'' of such systems, the leading order
asymptotics should behave similarly to the model described here.  On the
other hand, the subdominant asymptotics, for example the prefactor, would
require a different approach.

\smallskip

{\it Acknowledgments.\/} The authors are grateful to P.~Goldbart and
P.~Deift for stimulating conversations, and W.~Zhang for useful suggestions
on programming.  This work was supported in part by NSF Grants PHY-0651077
and PHY-0965015.

\bibliographystyle{ieeetr}
\bibliography{general}

\end{document}